\documentclass [12pt,a4paper]{article}
\usepackage{graphicx}
\usepackage{psfrag}
\usepackage{graphics}
\usepackage{bm}
\usepackage{color}
\usepackage{verbatim,color,ulem}

\newcommand{\gsim}{\raise.3ex\hbox{$>$\kern-.75em\lower1ex\hbox{$\sim$}}}
\newcommand{\lsim}{\raise.3ex\hbox{$<$\kern-.75em\lower1ex\hbox{$\sim$}}}

\begin{document}

\begin{center}
{\Large\textbf{Variable Modified Newtonian mechanics}}
\\[0.8\baselineskip]
{\Large\textbf{VIII. Ultra Faint Dwarf in Milky Way}}
\\[\baselineskip]
 {\large {James C.C.\ Wong}}
 \\[0.5\baselineskip]
 Department of Electrical Engineering,
     City University of Hong Kong. H.K.
\\[0.7\baselineskip]
\end{center}

\date{\today}
\begin{abstract}
Velocity dispersions of Ultra faint dwarf (UFD) galaxies are observed to deviate significantly from the Baryon Tully Fisher Relation of the massive galaxies (BTFR). We propose a solution within  Variable Modified Newtonian dynamics model, which could explain this deviation.
\end{abstract}


\section{Introduction}
Significant Mass discrepancies are observed at astrophysics and cosmology scales, where the common paradigm used to account for these discrepanies is cold dark particles. At solar system scale, gravitational observations so far do not require introduction of non-baryonic matter \cite{tremaine}-\cite{banik}. At galactic scales, the observed Radial Acceleration Relation (RAR) \cite{mcgaugh2004}-\cite{lelli} deviates significantly from Newtonian gravity predictions. To match the RAR observations, the $\Lambda CDM$ model requires fine tuning of baryonic physics \cite{mcgaugh2000}-\cite{abadi}. The observed RAR can also be well explained by a simple additional gravitational acceleration $\ddot{r}=\sqrt{g_Na_0}$ (called MOND acceleration which is proposed by Milgrom \cite{milgrom}) where $g_N$ is the Newtonian gravitational acceleration and $a_0$ is a fixed acceleration scale $a_0=1.2\times 10^{-10}m/s^2$. This acceleration  becomes dominant at $g_N\ll a_0$.  Observationally, the success of this MOND ansatz is limited to the mass range $10^{6.5}M_{\odot}-10^{11}M_{\odot}$ \cite{mcgaugh2010}. For Bright Cluster Galaxy (BCG) with mass $10^{11.5}M_{\odot}$ the MOND acceleration required to fit observation takes on a larger value $7a_0-20a_0$ \cite{tian}. In fact, BCGs with mass upto $10^{14}M_{\odot}$ are shown to follow a distinctly different RAR \cite{tian2} from the RAR of field galaxies of \cite{mcgaugh2004}.  Inside a galactic cluster, the difficulties appear at multiple scales \cite{aguirre}-\cite{li2023}. For the small mass dwarf spheroidals $M\leq 10^6M_{\odot}$, it is found that most dwarfs with large masses $M\sim 10^6M_{\odot}$ can still be explained by MOND \cite{sanders2021}-\cite{loeb2021}, while the smaller mass $M<10^5M_{\odot}$ (UFD) have velocity dispersions significantly higher than the MOND predictions. The tidal disruption proposal in \cite{mcgaugh2010} remains to be fully tested \cite{loeb2021}. The MOND ansatz requires additional non-CDM component in the cosmic background to account for the very large scale gravitational effect observed in the CMB power spectrum \cite{mcgaugh2014}, the formation of large scale structure \cite{nusser2002} and the Hubble parameter value over a range of redshifts by DESI BAO data \cite{karim}.  The limitations of the MOND program at small scales and large scales can play the role of signposts toward finding a FUNDAMOND model advocated by Milgrom \cite{milgrom2025} which should be a relativistic extension with a non-canonical $a_0$ plus an additional non-matter component in the cosmic background. To explain the UDF deviations from BTFR in CDM model, external effects from reionisation, supernovae explosion and various ways of interaction with the host galaxy are proposed as a primary process to strip off baryons from the UFD, which remain to be confirmed.  Moreover, at very high redshifts, the  observed massive galaxies and super massisve black holes suggest that the missing mass problem is more severe at very early times than can be easily accommodated by $\Lambda CDM$ \cite{mcgaugh2024}.  Early and late time cosmological measurement discrepancies, namely the Hubble tension \cite{riess}-\cite{valentino}, the $\sigma_8$ tension \cite{bohringer}-\cite{einasto} and the phantom crossing of dark energy from DESI DR2 \cite{karim}-\cite{ozulker}, if unresolved would necessitate a need to modify the $\Lambda CDM$ model.  Nonetheless, the recent ACT CMB data anaylsis \cite{louis} appears to rule out a large number of interesting models (including axion dark matter and interacting dark matter) currently under active investigation.  It is also reported recently that no evidence of the Sterile neutrino is found after a 10 year search \cite{microboone}. As it stands, any model that is not ruled out by the large amount of observational data so far would be worth serious attention. These models include Relativistic MOND \cite{skordis1}-\cite{skordis} and MOG \cite{moffat2}, although MOG's description of Matter Power Spectrum remains disputed \cite{dodelson}.
\\\\ 
In \cite{wong}, we propose a new solution of Einstein equation for multiple asymptotically dominant sources to explain missing source of gravity, in which the free fall velocities (not accelerations) due to asymptotically dominant sources are additive. For a central mass in an expanding cosmic background, this naturally introduces a MOND-like non-Newtonian acceleration with $a_0=\frac{1}{2}H(z)^2r$ which changes the baryon overdensity evolution equation. An overdensity from recombination grows much faster than its Newtonian perturbation theory counterpart and can turn around at a much higher redshift $z_{ta}$. As the overdensity collisionlessly collapses via a violent relaxation, in its virialised sphere, $H(z)$ in the non-Newtonian potential is kept at the value at $H(z_{ta})$.  This non-Newtonian potential can play the same role as an invisible matter halo potential.  For a spherical galaxy of $10^{10.5}M_{\odot}$, the corresponding MOND acceleration $a_0^{VM}$ matches the observed MOND acceleration at the correct radius \cite{wong2}. For the large galaxy $10^{11.5}M_{\odot}$, we recover $a_0^{VM}\sim 10a_0$ which is the observed value. In the Milky Way, the predicted rotational curve at $17kpc-28kpc$ matches Gaia DR2 observations \cite{wong6}. This paradigm of additive free fall velocities due to dominant background components leading to an additonal energy density which plays the role of dark matter and explains the CMB angular power spectrum \cite{wong3}, the $\sigma_8$ problem \cite{wong4}, the high redshift supermassive BH \cite{wong5}. Keeping the observed sound horizon angular scale from CMB data, with a cosmological constant, this formulation leads to a Hubble parameter that matches the DESI DR2 data with a Hubble constant at $H_0=72.8 km/s/Mpc$ \cite{wong5}. 
\\\\
In the small mass UFDs with $M\leq 10^5M_{\odot}$ considered in McGaugh-Wolf \cite{mcgaugh2010}, there a noticable correlation that the fainter, the lower metallicity  the UFD, the stronger its velocity dispersions deviate from its BTFR prediction. This missing mass problem is at its most severe for UFDs at its smallest mass end $M\leq 10^4M_{\odot}$ where the missing mass is 10 times to over 100 times of the baryonic mass. Both CDM model and MOND subscribe to external effects such as tidal disruptions which lead to a system inequilibrium, as the primary cause of this deviation. 
\\\\
In this work we propose an alternative mechanism for these deviations from BTFR. Within the VMOND formalism, there is another tidal effect scenario in which  due to external source of angular momentum, an overdensity  growth is stopped at very high redshift and decouples from the expanding background. In this case, the velocity dispersions in the overdensity can be very high and compatible with observations.
\\\\
In section 2, for completeness, we recapitulate the essential features of model and highlights its application to the rotational velocity (velocity dispersions) for different galaxies we study so far. In section 3, for two small mass UFD where velocity dispersions are extremely high compared to the MOND prediction, we  propose a scenario within VMOND, in which these high velocity dispersions can be produced. Section 4 is a summary and discussion.
\section{The Model}
\noindent In \cite{wong}, we notice that in the Tolman-Lema$\hat{i}$tre spherical symmetric metric solution to Einstein equation, specifying the free falling velocity of a particle specifies the metric. The particle free falling velocity around a central point mass is, according to Newton's law,
\begin{equation}
\dot{r} =-\sqrt{\frac{2GM}{r} }
\end{equation}
where $G$, $M$ and $r$ are the Newton's constant, the central point mass and the radial
distance respectively. From this free falling speed one obtains the Schwarzschild  Lema$\hat{i}$tre metric with coordinate time $\tau$, comoving distance $\varrho$ ($d\Omega^2=d\theta^2+sin^2\theta d\varphi^2$)
\begin{equation}
ds^2=c^2d\tau^2-\frac{2GM}{c^2r}d\varrho^2-r^2d\Omega^2.
\end{equation}
This metric can be transformed to the more familiar {\it Schwarzschild metric} in curvature coordinates.
\\\\
If we specify the free falling speed of a particle in a cosmological background with Hubble parameter $H(z)$ at redshift $z$, according to Hubble's Law,
\begin{equation}
\dot{r}=H(z) r,\:\:\: H(z) =\frac{1}{a(\tau)} \frac{d a(\tau)}{d\tau},
\end{equation}
where $a(\tau)$ is the scale factor at $z$ and we obtain the Friedmann-Lema$\hat{i}$tre metric
\begin{equation}
ds^2=c^2d\tau^2-a^2 d\varrho^2-r^2d\Omega^2.
\label{frw}
\end{equation}
For a central point mass in an expanding background, in \cite{wong} we find a new metric solution having the form 
\begin{equation}
ds^2=c^2d\tau^2-\frac{2GMa^3}{c^2r}d\varrho^2-r^2d\Omega^2,
\label{LT1}
\end{equation}
which, for a weak gravitational potential, is equivalent to the perturbed FRW metric in Conformal Newtonian gauge. Here the underlying free fall velocity is given by
\begin{equation}
\dot{r}= H(z)r-\sqrt{\frac{2GM}{r}}.
\label{dotr1}
\end{equation}
Eq.(\ref{dotr1}) depicts that the particle will follow the Hubble law at large distances, but at small distances it will follow a Newtonian free falling velocity. This equation differs from the $\dot{r}$ choice in Schwarzschild de-Sitter metric,
\begin{equation}
\dot{r}^2=\frac{2GM}{r}+\frac{\Lambda c^2 }{3}r^2,
\end{equation}
where $\Lambda$ is the cosmological constant.  At slow speeds in radial direction, Eq.(\ref{dotr1}) is formally the same as the equation of motion in Newtonian perturbation theory
\begin{equation}
\dot{r}=H(z)r+v_p,\:\:\:\:\ddot{r} =\frac{\ddot{a}}{a}r+\dot{v}_p,
\label{NPT}
\end{equation}
where $v_p$ is the peculiar velocity of the particle. Eq.(\ref{NPT}) is the base equation used to obtain the overdensity evolution in the Newtonian perturbation theory \cite{mukhanov}.
Contrary to the Newtonian perturbation theory, due to the radius dependence in $v_p=-\sqrt{2GM/r}$, Eq.(\ref{dotr1}) in fact leads to the acceleration equation 
\begin{equation}
\frac{d \dot{r}}{dt}=\dot{r} \bigg(\sqrt{\frac{GM}{2r^3}}+H\bigg)+r\dot{H},
\label{ddotr0}
\end{equation}
\begin{equation}
\ddot{r}=-\frac{GM}{r^2} -\sqrt{\frac{H^2r}{2}}\sqrt{\frac{GM}{r^2}}+\frac{\ddot{a}}{a}r =-\sqrt{\frac{GM}{r^2}a_0^{VM}} +\frac{\ddot{a}} {a}r,.
\label{nNewton1}
\end{equation}
We term the sum of the Newtonian and the non-Newtonian gravitational acceleration  the "VMOND" acceleration, where $a_0^{VM}$ corresponds to the $a_0$ in the canonical MOND paradigm at large distance limit. In this model, there is no interpolation function between Newtonian acceleration to deep MOND acceleration regime to be specified.
The radial acceleration including angular momentum is given as
\begin{equation}
\ddot{r}=  \frac{h^2}{r^3} -\frac{GM}{r^2}-\sqrt{\frac{1}{2}H^2(r)r}\sqrt{\frac{GM} {r^2} } +\frac{\ddot{a}} {a}r,
\label{vmond}
\end{equation}
where $h$ is the angular momentum per unit mass.  This is the geodesic equation (29) given in Baker \cite{baker}.
To obtain the geodesic equation Eq.(\ref{vmond}), one can differentiate w.r.t $\tau$ the following 
the slow speed energy equation
\begin{equation}
\frac{1}{2}\dot{r}^2+\frac{h^2}{2r^2} =E+\frac{1}{2}\bigg(\sqrt{\frac{2GM}{r}}-H(z)r\bigg)^2,
\label{potl44}
\end{equation}
where $E$ is the energy of the particle. 
%
The distance from the centre
\begin{equation}
r_{ta}=\bigg(\frac{2GM}{H^2}\bigg)^{1/3}
\label{rh}
\end{equation}
describes a turnaround radius of the particle. For a central solar mass at late time, we have $r_{ta} \sim 10^{7}AU$, but the observed solar system scale is $10^{5} AU$. This means that within solar system scale, the dominant gravitational acceleration remains Newtonian. This is observationally confirmed at Oort cloud scale by \cite{tremaine}.
\subsection{Density perturbation evolution}
The evolution of a baryon overdensity from recombination is presented in the first paper of this series \cite{wong}, here we recall some of the main features which will be relevant for our study.
\\\\
The Hubble parameter at redshift $z$ where radiation is negligible is given by adding the Hubble parameter $H_b$ due to baryons and Hubble parameter $H_{\Lambda}$ due to the cosmological constant,
$H(z)=H_b+H_{\Lambda}$. The Friedmann Equation takes the form,
\begin{equation}
H^2(z) =\frac{8 \pi G}{3}\rho_H=\frac{8\pi G}{3} \bigg(\rho_b+2\sqrt{\rho_b \rho_{\Lambda}}+\rho_{\Lambda}\bigg)
\label{H22}
\end{equation}
\begin{equation}
H^2(z) =H_0^2 \bigg(\Omega_b(1+z)^3+2\sqrt{\Omega_b \Omega_{\Lambda}}(1+z)^{3/2}+\Omega_{\Lambda}\bigg),\:\:\:H^2_0=\frac{8 \pi G}{3}\rho_c,
\label{H22}
\end{equation}
where $\rho_b$ and $\rho_{\Lambda}$ are cosmological background densities of baryonic matter and dark energy respectively. $H_0$  and $\rho_c$ are the Hubble parameter and the critical density of the present epoch. $\Omega_b$ and $\Omega_{\Lambda}$ are the density parameter for baryons and $\Lambda$ respectively. When including the radiation density, this Hubble parameter can explain the dark matter potential at early times \cite{wong3} and has a matter dominant epoch at $z=600-1.67$ and therefore performs in this epoch effectively in the same way as the Friedmann-Lema$\hat{i}$tre-Robertson-Walker (FLRW) cosmology.
\\\\
Soon after recombination, the cosmological background is in the baryon matter dominant epoch with mean density 
\begin{equation}
\rho_b=\frac{1}{6\pi G\tau^2}.
\end{equation}
The baryon density perturbation $\delta$ is specified by 
\begin{equation}
\delta = \frac{ \rho-\rho_b} {\rho_b}=\frac{\delta\rho_b}{\rho_b},
\label{delta}
\end{equation}
where $\rho$ is the total mass density of the overdense region inside radius $r$. The baryon matter overdensity is given by $ \delta \rho_b$.
\\\\
In the matter dominant era,  the non-Newtonian acceleration in Eq.(\ref{nNewton1}) corresponds to the "dynamical" mass density $\delta^{1/2}\rho_b$. From \cite{wong} the  effective overdensity evolution equation becomes
\begin{equation}
\frac{\partial^2}{\partial \tau^2}\Delta(\delta) +2H\frac{\partial}{\partial \tau} \Delta(\delta) = 4\pi G \rho_b \Delta(\delta).
\label{deltadot22}
\end{equation}
where $\Delta(\delta) =\delta+\delta^{1/2}$.
At the matter dominant epoch where $H=2/(3\tau)$, Eq. (\ref{deltadot22}) has the growth solution
\begin{equation}
\Delta(\delta)=\delta +\sqrt{\delta} \propto  \tau^{2/3} \propto a(\tau);\:\:a(\tau)\propto \frac{1}{1+z}.
\label{Delta1}
\end{equation}
In the VMOND paradigm, at $\delta\ll1$, we obtain a new $\delta$ growth rate $\delta \propto \tau^{4/3} \propto a(\tau)^2$. In the MOND paradigm, the same overdensity growth rate $\delta\propto a^2$ is used in large scale structure simulation \cite {nusser2002} and in elliptical galaxy formation simulation \cite{sanders2007}. 
\\\\
From an initial overdensity $\delta_{int}$ at recombination, we can use $\delta+\delta^{1/2}\propto a(\tau)$ to calculate the overdensity $\delta$ at redshift $z<1080$ by
\begin{equation}
\delta+\sqrt{\delta}=\bigg(\delta_{int}+\sqrt{\delta_{int}}\bigg) \bigg(\frac{1081} {1+z}\bigg)=\frac{A_0}{1+z},\:\:\: A_0 =\bigg(\delta_{int}+\sqrt{\delta_{int}}\bigg)1081.
\label{dpsd}
\end{equation}
where $A_0$ is fixed by the value of $\delta_{int}$. In a direct collapse model, one can estimate a massive galaxy's initial overdensity from the CMB power spectrum at recombination. 
In a cubic box approximation, at characteristic box size scale $R$ the overdemsity behavies as $\delta (R) =A_s R^{-S},\:\:0<S<3$ \cite{binney} (pp720-721),\cite{nusser2005}. The range value of $A_s$ can be estimated from \cite{verde1}.
\\\\
In \cite{wong2}, we choose $\delta_{int}=2.8\times 10^{-3}$ ($A_0=60$) to develop into a $10^{10.5}M_{\odot}$ elliptical galaxy. This choice is compatible with Sanders \cite{sanders2007} and Bromm \cite{bromm2}. 
\\\\
For a massive galaxy, we assume that the external angular momentum does not affect the evolution of the overdensity which will reach its turnaround at redshift $z_{ta}$ where $\delta=1$. From Eq.(\ref{dpsd}),
\begin{equation}
1+z_{ta}=\frac{A_0}{2}=30, \:\:\: z_{ta}=29.
\label{ta}
\end{equation}
After the turnaround, in \cite{wong2} we find an approximate particle free fall time $\tau_{ff} =\tau_{ta}$ which is also the violent relaxation time scale \cite{lyndenbell}. In a violent relaxation scenario,  virialisation $z_{vr}$ corresponding to different stage of virialisation is given by \cite{wong2}. (In the dark matter halo collapse model \cite{bromm2}, their virialised time is taken at $\tau=\tau_{ta}+\tau_{ff}$ where a tight (but messy) central configuration could occur.) This virialised redshift  $z_{vr}$ can be shown to  occur at $z_{vr}=17.9$ in \cite{wong2}, with a Quasi-Stationary-State (time independent state) occurred at
\begin{equation}
z_{vr}=\frac{1+z_{ta}}{3.39}-1=7.85
\end{equation}
and an effectively complete virialisation occurs at
\begin{equation}
z_{vr}=\frac{1+z_{ta}}{6.0}-1=3.88.
\end{equation} 
The recent observation of a mature massive spiral galaxy at $z\sim 4$ \cite{jain} fits in naturally in this paradigm.
 %
%
\subsection{Virialisation at high redshift and $a_0^{VM}$}
After an overdense cloud turnarounds and collapses gravitationally at high redshift to reach a central core, we expect that the cloud primarily relaxes through a "violent relaxation" similar to what is found in the three-dimensional dissipationless collapse simulation in a MOND potential by Nipoti et al. \cite{nipoti} and reaches a meta-equilibrium first and eventually fully virialises. 
\\\\
The central mass upto a shell at $r$, with a power-law density $\rho(r)=\bar{\rho}r^{-S}$ ($\bar{\rho}$ is constant and $S>0$) is given by
\begin{equation}
M(R)=4 \pi \int_0^{R} r^2 \rho(r) dr =4 \pi \bar{\rho}\int_0^{R} r^{-S} r^2dr=\frac{4 \pi \bar{\rho}} {(3-S)} R^{(3-S)}.
\label{mt}
\end{equation}
Given the time averaged Kinetic energy $K=\frac{1}{2}M V^2$ (here $V$ is the average rotational speed) and average potential energy $\Psi$, the virialised relation is given by $2K+\Psi=\bar{E}$, with the average energy $\bar{E}\rightarrow 0$ at equilibrium.
\\\\
The virial potential energy is given by
\begin{equation}
\Psi=-4\pi\int_0^{R} drr^3 \rho(r) \nabla \cdot \Phi,
\end{equation}
where $\Phi$ is the local potential. The virial potential consists of both a Newtonian and a non-Newtonian potential
\begin{equation}
\Psi=-4\pi \int_0^R dr r^3 \bar{\rho}r^{-S} \bigg(\frac{GM(r)}{r^2}+  H(z_{ta})\sqrt{ \frac{GM(r)}{2r}}\bigg),
\label{psi}
\end{equation}
\begin{equation}
=-C_1(S) \frac{GM^2}{R}-C_2(S)MH(z_{ta})\sqrt{\frac{GMR}{2}}.
\end{equation}
\begin{equation}
 C_1(S)=\bigg(\frac{3-S}{5-2S}\bigg) ;\:\:\:C_2(s)= \bigg(\frac{3-S}{5-\frac{3}{2}S}\bigg).
\end{equation}
The MOND-like virial potential comes through a collisionless relaxation and virialisation process, we assume that there is no loss of potential energy corresponding to this term and keep the redshift "$z_{ta}$" in the evaluation of this (redshift dependent) virialised potential. This primeval turnaround redshift $z_{ta}$ value is crucial in providing the strong non-Newtonian gravity at galactic scales as we shall find below. The acceleration due to the cosmic background which cannot be virialised is usually treated as non-participating according to the "Jeans Swindle" which is also used in MOND simulations \cite{nusser2005}, \cite{sanders2007}. In our paradigm, the effect of the cosmic background acceleration to a particle in virialised potential remains. When the virialised system is considered at a redshift and radius much lower than $z_{ta}$ and $r_{ta}$, the overall effect due to the background accleration is negligible for our purpose.
It is worth noting that if the non-Newtonian potential is treated as a dark halo (potential), it can lead to an apparent {\it dark halo-visible matter coupling relation} observed in \cite{salucci}, but this potential is not easily distinguishable from collisionless cold dark particle halo based on gravitational effects alone.
\\\\
For a stationary orbit where the kinetic energy is dominated by an averaged rotational velocity $V^2$, (for system with line of sight velocity dispersions $\sigma$, we follow \cite{mcgaugh2010} to take $V^2=3\sigma^2$) we have 
\begin{equation}
V^2(R)=C_1(S)\frac{GM}{R}\bigg(1+\frac{C_2(S)} {C_1 (S) }\sqrt{\frac{H^2(z_{ta})R^3}{2GM}}\bigg) =C_1(S)\frac{GM}{R}\bigg(1+ \frac{C_2(S)} {C_1 (S) }\sqrt{\frac{\rho_H(z_{ta})}{\rho_b(R)}}\bigg),
\label{virialeq1}
\end{equation}
where $\rho_H(z_{ta})$ is the cosmic background density at $z_{ta}$. At short distnce $R$, the baryon density $\rho_b(R)$ is high such that $\rho_b(R)\gg \rho_H(z_{ta})$, we have the nearly Newtonian behaviour from Eq.(\ref{virialeq1}). As a mass shell at turnaround collapses to a shell at radius $R$, baryonic mass conservation leads to 
\begin{equation}
M=\frac{4\pi}{3} \rho_b(R)R^3=\frac{4\pi}{3}\rho(z_{ta}) r_{ta}^3.
\end{equation}
This equation leads to
\begin{equation}
\frac{\rho_b(R)}{\rho(z_{ta})} =n^3,\:\:\:\: n=\frac{ r_{ta}}{R}.
\end{equation}
which relates the rotational speed at $R$ with the turnaround radius $r_{ta}$.
We can also rewrite Eq.(\ref{virialeq1}) to emphasize the large radius behaviour as
\begin{equation}
V^4(R)= C_2(S)^2GM \bigg(\frac{1}{2}H(z_{ta})^2R\bigg) \bigg(1+\frac{C_1(S)}{C_2(S)}\sqrt{\frac{\rho_b(R)}{\rho_H(z_{ta})}}\bigg)^2 =GMa_0^{VM}(z_{ta}, R).
\label{vsigma11}
\end{equation}
At large $R$, $\rho_b(R)/\rho_H(z_{ta})$ approaches a number close to unity.
Eq.(\ref{vsigma11}) with an emphasis on the far R.H.S. can be recognised as the Faber-Jackson relation or Tully-Fisher relation with a variable $a_0^{VM}(z_{ta}, R)$. 
\\\\
In \cite{wong2}, we find that the virialised potential of a sphere has a Newtonian  dominant central region around its half mass radius and a MOND like region at large radus, this is compatible with the findings in Durazo et al. \cite{durazo} over multiple orders of magnitude in galactic mass. 
\subsection{A summary of our findings of $a_0^{VM}$ for different galactic systems}
Using $\delta_{int}=2.8\times 10^3$, we find the followings
\\\\
1) For large distances behaviour, we choose, from \cite{suess}, a large star forming elliptical galaxy with mass $M_{*}=10^{10.5} M_{\odot}$ with $R_e\sim 3kpc$ and data is given out to $r_{18}=6R_e=18kpc$. We find that $r_{ta}=39.2kpc$ and 
\begin{equation}
a_0^{VM}(6R_e)=1.05 a_0,
\end{equation} 
in remarkable qualitative agreement with the $a_0$ of MOND analysis. There is a galaxy with similar mass given in \cite{durazo} and \cite{milsanders}, where the velocity dispersions can be given upto to $R=7R_e$. For this cases, the same calculation yields $a_0^{VM}(7R_e)=0.86a_0$. If we extrapolate the calculation to $R=r_{ta}=13R_e$, we will obtain $a_0^{VM}(13R_e)=0.51 a_0$. This result means that at very large distances, Eq.(\ref{vsigma11}) persists with a weakly decreasing $V\propto a_0^{1/4}$, although observational data extended to such a large radius is rare.
\\\\
2) For a larger galaxy at the scale of BCG with mass $10^{11.5}M_{\odot}$, which has small $R_e =1-4kpc$. For the same initial overdensity, we have $z_{ta}=29$ as before, the turnaround radius is larger at $r_{ta}=2.15\times 39 kpc$, we can evaluate $a_0^{VM}$ around $18kpc$ as before and obtain
\begin{equation}
a_0^{VM}(18kpc)=7.37a_0,\:\:\:a_0^{VM}(15kpc) =9.94a_0.
\end{equation}
This is consistent with the observed MOND acceleration is $a_0^{BCG}=7.5a_0-20a_0$ reported in \cite{tian} for a sample of 54 MaNGA BCGs.
\\\\
3) Gaia DR3 provides a highly precise data of MW rotational curve from $6kpc-28kpc$, which is shown to be problematic for MOND \cite{chan}-\cite{coquery} and MOG \cite{moffat}. In \cite{wong3}, we show that by flattening a virialised sphere into a virialised Mestel disk \cite{mestel}, we obtain rotational curve prediction which matches observed data very closely.
\\\\
4) Kinematics of Tidal Dwarf Galaxies (TDG) which are formed by debris at late time $z\simeq 0$ is found to be compatible with Newtonian gravity with no Dark matter \cite{lelli3}. It also poses a challenge to the canonical MOND paradigm. In the VMOND paradigm, here the turnaround redshift $z_{ta} \simeq 0$, the non-Newtonain potential in Eq.(\ref{psi}) is at $O(10^{-3})$ times the Newtonian potential and the velocity dispersions therefore should be dominated by Newtonian acceleration.
\\\\
5) Some Milky Way low surface brightness spheroidal galaxies are found to have large velocity dispersions but remain compatible with MOND for some galaxies \cite{sanders2021}.  We consider the galaxy Carina in \cite{sanders2021} where $M=2.4\times 10^6M_{\odot}$, the effective radius $r_{eff}=0.369\: kpc$, the line of sight observed velocity dispersions is averaged around $6.6\: km/s$. We assume that Carina is a small part of the large overdensity which formed the Milky Way and therefore having the same initial overdensity. This small cloud also decouples from the cosmic background at $z_{ta}=29$ at sufficiently high systematic angular momentum w.r.t to the galactic centre, which keeps the gas cloud to stay in a large stable orbit around the Milky Way centre. This gas subsequently gravitationally collapses onto its own mass centre which resides in the large orbit. We assume that this small cloud gravitational collapse process behaves similarly to the collapse process as the large central galaxy discussed above. For the mass of Carina, the turnaround radiius is $2.67 kpc$. Consider $R=0.5kpc$, $n=5.34$,
the velocity dispersion 
\begin{equation}
\sigma(0.5kpc)=\bigg(\frac{1}{9} GM a_0^{VM}\bigg)^{1/4}= 5.9 km/s.
\end{equation}
This is within the line of sight observation scatter of $\sigma_{los} = 6.6\pm 2.6\: km/s$ in \cite{sanders2021}.
\section{Implication for allowable rotational velocities for Coma Berenices.}
So far our base model is a Monolithic overdensity turnarounds, gravitational relaxes and virialises model. The rotational velocity from Eq.(\ref{virialeq1}) and Eq.(\ref{vsigma11}) can be rewritten as
\begin{equation}
V^4 =GM\bigg(\frac{1}{8} H^2_0\Omega_b\bigg) (1+z_{ta})^3 r (1+2n^{3/2})^2
\label{v4gm}
\end{equation}
Consider the dwarf galaxy Coma Berenices (CB) data in \cite{mcgaugh2010}, where its half-light radius $r_{1/2}=100pc$ and mass within $r_{1/2}$ is $M(r_{1/2})=10^{4} M_{\odot}$. $\Omega_b=0.044$, $\rho_c=10^{-26}kg/m^{_3}$, $V(r_{1/2})=8\pm 1.4 km/s$. From Eq.(\ref{v4gm}).
\begin{equation}
V^4(r_{1/2}) =1.29\times 10^{5}m^4s^{-4} (1+z_{ta})^3 (1+2n^{3/2})^2.
\label{v4}
\end{equation}
The baryon density of CB at $r_{1/2}$ is $\rho_b(r_{1/2})=1.76\times 10^{-22}kg/m^3$. For gravitational collapse to proceed, the baryonic density at turnaround must be smaller than $\rho_b(r_{1/2})$. Using Eq.(\ref{Delta1}), one finds that the maximum overdensity at recombination is $\delta_{int}=1.28\times 10^{-2}$, where $1+z_{ta}=73.68$, $r_{ta}/(r_{1/2})=n=1$, we obtain from Eq.(\ref{v4}) 
\begin{equation}
V(r_{1/2})=0.826km/s.
\end{equation}
To obtain minimum initial overdensity, we take the mean CMB baseline baryonic overdensity at recombination which is $\delta_{int} =3\times 10^{-5}$, where $1+z_{ta}=2.95$, $r_{ta}/(r_{1/2})=11.6$. Although this scenario is unlikely to explain UFD observed at higher redshift, for discussion purposes we obtain
\begin{equation}
V(r_{1/2})=0.676 km/s.
\end{equation}
These rotational velocities are similar to the MOND estimate given in \cite{loeb2021}.
Given the observed $V(r_{1/2})=8\pm 1.4km/s$, we note that the successful model for massive galaxy formation, based on overdensity turnarounds, violently relaxes and virialise, is unlikely to explain the high velocity disperson of CB. Tidal disruptions due to the host galaxy could be advocated to explain this deviation as in CDM and MOND \cite{mcgaugh2010}, but here we explore a different scenario of tidal disruption in VMOND that could produce this velocity discrepancy.
\subsection {Effect of a large Angular momentum at very high redshifts}
The acceleration of a particle on an overdensity mass shell Eq.(\ref{vmond}) can be written in terms of densities in matter dominant epoch is
\begin{equation}
\ddot{r}=\frac{h^2}{r^3}-\bigg(1+\Delta \bigg)\frac{4\pi G\rho_b(z) r}{3},\:\:\Delta =\delta+\sqrt{\delta}.
\label{acceqn}
\end{equation}
In Newtonian perturbation theory, where $h^2=0$ is assumed for spherical overdensity such that Eq.(\ref{acceqn}) is essential for the overdensity evolution $\Delta=\delta \propto a(\tau)$.
In practice, during the overdensity growth, the overdense region can receive an angular momentum due to tidal fields from its neighbours. A common assumption is that angular momentum behaves as a small perturbation and does not affect the overdensity growth. 
Specifically, for physical distance $r$, one has $r=a(\tau) x(q,\tau)$, $\delta(x,\tau)=Ka(\tau)\delta(x)$, $K$ being constant, the specific angular momentum due to tidal fields is given by, \cite{peebles}-\cite{white1},
\begin{equation}
h(\tau)=\frac{K\rho_b a^3 \dot{a}}{M} \int_{V_L} d^3q\: q \times v(q,\tau)
\end{equation}
where $V_L$ is the volume of the overdensity, $v$ is the velocity at distance $q$ from the mass centre. 
\\\\
In the presence of a heavy Void, Casuso and Burkert \cite{casuso} -\cite{burkert} propose that 
\begin{equation}
h(\tau)=\frac{K\rho_b a^3 \dot{a}}{M_G} \int_{V_L} d^3q\: q \times (-2\Omega \times \textbf{i} \sqrt{\frac{GM_Gd}{q^2}})
\label{ht}
\end{equation}
where $M_G$ is the galaxy mass, $\Omega \propto M_T^{2/3}$ is the mean angular velocity of the Void, $M_T$ is the mass of the Void and $d$ is the distance travelled by the gas during a gravitational collapse and $\textbf{i}$ is the unit vector along the rotating velocity of the Void. One obtains angular momentum with unit vector $\textbf{n}$, in the direction of the centre of the rotating Void, such that 
\begin{equation}
h(\tau) \propto M_T^{2/3}(GM_Gr)^{1/2}.
\label{h}
\end{equation}
In this scenario, the angular momentum transfer from the Void causes the break-away of the overdensity. 
\\\\
A typical Void mass is $M_T \propto 10^{15} M_{\odot}$ \cite{casuso}-\cite{burkert}. For a small mass overdensity $M\leq 10^4M_{\odot}$, from Eq.(\ref{acceqn}) the required angular momentum to cause an overdensity to stop growing and break away from cosmic background is 
\begin{equation}
\dot{\varphi}^2= \frac{h^2}{r_i^4}=\bigg(\delta_i+\sqrt{\delta_i} \bigg)\frac{4\pi G \rho_b(z_i) }{3},\:\:\:h^2= GM_Gr_i \bigg(1+\frac{1}{\sqrt{\delta_i}}\bigg),
\label{h2}
\end{equation}
which is easily attainable given the large mass $M_T$ of the Void from Eq.(\ref{h}). Apart from a Void, after recombination, VMOND describes an environment in which the relative accelerations amongst massive overdensities take on a much larger value, similar to the effect of MOND on El Gordo cluster \cite{kroupa}. This large relative velocities can also lead to a larger tidal effect between neighbouring overdensities.
\\\\
In this work we assume the condition in Eq.(\ref{h2}) is attained for a small uniform overdensity mass shell at some distance $r_i$, $z_i$ and $\delta_i$. The overdensity mass in the spherical mass shell is given by
\begin{equation}
M(r_i)=\frac{4\pi}{3}\delta_i\rho_b(z_i)r_i^3=\frac{4\pi}{3}\delta_i\rho_c \Omega_b (1+z_i)^3r_i^3
\label{mri}
\end{equation}
Since this overdensity will contract in radius from $r_i$ to  $r_{1/2}$, where $r_i=n r_{1/2}$, while $M(r_{1/2})=M(r_i)$, from Eq.(\ref{mri}) we obtain
\begin{equation}
n\delta_i^{1/3}(1+z_i) = \bigg(\frac{3M(r_{1/2})}{ 4 \pi\Omega_b \rho_c r_{1/2}^3}\bigg)^{1/3}.
\label{ndeltaz}
\end{equation}
\subsection{The rotating velocity due to freezed overdenstiy}
We start with the lowest realistic overdensity $\delta_{int}=3\times 10^{-5}$ at recombination, we obtain
\begin{equation}
\bigg(\delta_{int}+\delta_{int}^{1/2}\bigg) \times 1080 = 6.22=\bigg(\delta_i+\delta_i^{1/2}\bigg) (1+z_i).
\label{ndeltaz3}
\end{equation}
From Eq.(\ref{h2}), the rotational velocity at break-away radius $V_i$ takes the form
\begin{equation}
V_i^2=\frac{h^2}{r_i^2}=\frac{GM}{r_i} \bigg(1+\frac{1}{\sqrt{\delta_i}}\bigg).
\end{equation}
If there is a contraction $r_i=nr_{1/2}$, the rotational velocity at $r_{1/2}$ is 
\begin{equation}
V(r_{1/2})^2=n\frac{GM}{r_{1/2}}\bigg(1+\frac{1}{\sqrt{\delta_i}}\bigg).
\label{vndeltai}
\end{equation}
Consider the dwarf galaxy CB parameters, from Eq.(\ref{ndeltaz}) we obtain
\begin{equation}
n\delta_i^{1/3} (1+z_i)=76.2.
\label{ndeltaz1}
\end{equation}
 For small $\delta_i $, from Eq.(\ref{ndeltaz3}) we have $\delta_i^{1/2}(1+z_i)\simeq 6.22$. Together with Eq.(\ref{ndeltaz}) one obtains
\begin{equation}
n=12.24 \delta_i^{1/6},\:\:\:\: n^3(1+z_i)= 1.13\times 10^4.
\end{equation}
If the overdensity stops growing at $\delta_i=10^{-3}$, we have $n=3.87$, $1+z_i=195$ and 
\begin{equation}
V(r_{1/2})= 7.45km/s.
\end{equation}
We note that deviation from sphericality can lead to an increase of rotational speed upto $20\%$ \cite{mcgaugh1998}, which will bring the prediction in-line with observation.
To obtain generically a higher rotational velocity, we can take $\delta_i =6\times 10^{-4}$ and obtain
$n=3.55$, $1+z_i=252$ and 
\begin{equation}
V(r_{1/2})= 8.11km/s.
\end{equation}
Next, we consider Segue1, the smallest UFD in \cite{mcgaugh2010}, with $M(r_{1/2})=10^{2.78\pm 0.5}M_{\odot}$, $V(r_{1/2})=7.4\pm 2.0 km/s$, $r_{1/2}=38\pm 9pc$. We choose for our simple model $M(r_{1/2})=10^3M_{\odot}$, $r_{1/2}=40pc$. Repeat the above calculation we obtain
\begin{equation}
n=19.13 \delta_i^{1/6},\:\:\:\: n^3(1+z_i)= 4.36\times 10^4.
\end{equation}
If the overdensity stops growing at $\delta_i=6\times 10^{-4}$, we have $n=5.54$, $1+z_i=252$ and
\begin{equation}
V(r_{1/2})= 4.81km/s.
\end{equation}
Adding a factor of $20\%$ due to non-sphericality, we obtain $V(r_{1/2})=5.78km/s$ which reaches the lower limit of the observed $V(r_{1/2})$.
\\\\
If the overdensity stops growing at $\delta_i=10^{-4}$, we have $n=4.11$, $1+z_i=622$ and
\begin{equation}
V(r_{1/2})= 6.45km/s.
\end{equation}
Adding $20\%$ due to non sphericality, one obtains $V(r_{1/2})=7.73km/s$. Although mathematically viable, we note that the value of $z_i$ is probably too high in practice. There are two assumption modifications which can improve this result.
\\\\
Firstly, since $\delta_{int} =3\times 10^{-5}$ is a mean value, we can consider $\delta_{int} =10^{-5}$ instead. Then we have 
\begin{equation}
n=34.8\delta_i^{1/6},\:\:\:\: n^3(1+z_i)=1.45\times 10^5.
\end{equation}
For $\delta_i=6\times 10^{-4}$, we have $n=10.1$, $1+z_i=140$ and 
\begin{equation}
V(r_{1/2})=6.83 km/s.
\end{equation}
A $10\%$ increase due to non-sphericality will match the observed velocity value.
\\\\
So far, we assume that the angular momentum transfer to the small mass overdensity is incremental and smooth. In practice, the angular momentum transfer can be disruptive that all we need for model to work is that an overdensity stops growing at 
\begin{equation}
\frac{h^2}{r_i^3}\geq (\delta_i+\sqrt{\delta_i}) \frac{4\pi G}{3} \rho_b r.
\end{equation}
In this case, a $10\%$ increase of transfer momentum will suffice to obtain the observed rotational velocity.
\\\\
In summary, we show that when a small baseline overdensity from recombination encounters a large void, picks up sufficient angular momentum such that it stops growing at very high redshift, the resulting rotating velocity (velocity dispersions) can reach a very high (and non-Newtonian) value compatible with observations in UFDs. In the very high redshift non-Newtonian acceleration dominant regime with large voids or relatively fast moving clouds, the small mass of UFDs makes it easy for this scenario to occur. Also the smaller the overdensity mass, the more likely that the overdensity can stop growing at higher redshift where $\delta_i$ is smaller and therefore from Eq.(\ref{vndeltai}) leads to a larger rotational speed deviation from its equilibrium value at late time.
\subsection{The turnaround point of the largest mass shell}
When an overdensity with radius $r_i$ receives enough angular momentum and stops growing, the energy equation becomes
\begin{equation}
\frac{1}{2} \dot{r}^2= -E_0+\frac{h^2}{2r_i^2}-\frac{h^2}{2r^2} +\bigg(\frac{GM}{r}-H(z_i) \sqrt{2GMr}+\frac{1}{2}H(z)^2r^2\bigg),
\label{e}
\end{equation}
where the total energy is $E=\frac{h^2}{2r_i^2}-E_0$ and $E_0$ represents the additional radial velocity toward the mass centre which a particle on the mass shell picks up when bounced off the Void.
Here $H(z_i)$ remains fixed but $H(z)$ continues to follow the cosmic expansion. $M$ are now fixed by $\delta_i$ and $H(z_i)$. 
\\\\
Initially $H(z) \sim H(z_i)$, the energy equation behaves approximately as
\begin{equation}
\frac{1}{2}\dot{r}^2=-E_0 +\frac{h^2}{2r_i^2}-\frac{h^2}{2r^2}+\frac{\alpha G M}{r},\:\:\: 
\label{Eeqn}
\end{equation}
\begin{equation}
\alpha =\bigg(1-\sqrt{\frac{H^2(z_i)r^3}{2GM}}\bigg)^2=\bigg(1-\sqrt{\frac{1}{\delta_i n^3}}\bigg)^2.
\end{equation}
A particle on this mass shell can turnaround at some $r>r_i$ and returns to $r=r_i$ at $\alpha >0$.
\\\\
Consider a CB size protogalaxy with $\delta_i=10^{-3}$, $r_i=387pc \:(n=1)$, $\alpha\sim 10^3$ is very large and positive as $r$ approaches the mass centre. At the point $\delta_i n^3=1$ ($r=38.7pc$) we have $\alpha =0$ where the $\dot{r}^2<0$. This suggests that the particle should reach $\dot{r}=0$ at some $r_c>38.7 pc$. {\it This could put a constraint on the virialised matter density inside this mass shell.} 
\\\\
After this point, instead of following a Jeans Swindle approximation, we continue to work with a fixed $H(z_i)$ virialised non-Newtonian potential with a decreasing $H(z)$ due to matter cosmic background. As the particle moves away from the mass centre, the energy equation becomes
\begin{equation}
\frac{1}{2}\dot{r}^2=E-\frac{h^2}{2r^2}+\frac{GM}{r}\bigg(1-\frac{2}{\sqrt{\delta_i}}\sqrt{\frac{r^3}{r_i^3}} +\frac{1}{\delta_i}\frac{r^3}{r_i^3}\frac{(1+z)^3}{(1+z_i)^3} \bigg)
\label{e2}
\end{equation}
This equation is simplified at $r=r_i$
\begin{equation}
\frac{1}{2}\dot{r}_i^2=-E_0 +\frac{GM}{r_i}\bigg(1-63.2 +10^3\frac{(1+z)^3}{(1+z_i)^3} \bigg)
\label{Eeqn3}
\end{equation}
For $(1+z)/(1+z_i)>2..55$, the RHS of Eq.(\ref{Eeqn3}) is generically negative, and the particle is bounded at some  $r_f<r_i$. For $(1+z)/(1+z_i)=2$, RHS of Eq.(\ref{Eeqn3}) provides a minimum constraint on $E_0$ to stop the particle from outgoing at $r=r_i$. Once bounded, the subsequent orbits will also be bounded from outgoing at $r_i$ as $1+z$ decreases further.  (Using the free fall equation $\dot{r}=-\sqrt{2GM}{r}+H(z_i) r$ as a crude approximation \cite{wong2} for reference, we obtain $(1+z)/(1+z_i) =2.08$.)
\\\\ 
We also note that the baryon density inside $r_{1/2}$ of CB is $\rho(r_{1/2})=1.76\times 10^{-22} kg/m^3$. If for a choice of $\delta_i$ that the mass shell radius is bounded from below at $r=38.7pc$, its maximum uniform density is $\rho = 3\times 10^{-21}kg/m^3$ which is still lower than the hydrogen density $\rho =10^{-20}kg/m^3$ ($10^4\: atom/cm^3$) that is favourable for star formation. A lower bound on mass shell radius not far from $r_{1/2}$ could be a reason for a low overall star formation rate for the UFD..
\section{Summary}
The velocity dispersions of small UFD deviate significantly from BFTR for massive galaxies, that both CDM and MOND at equilibrium do not predict. Explanations are mainly due to different forms of external sources of disruptions. In our paradigm, using monolithic gravitational collapse model, we find that the predicted velocity dispersion of an UFD (Coma Ber.) is constrained close to the MOND prediction but is far from the observed value. We notice that at the very high redshift epoch, the inter-galactic scale non-Newtonian gravitational acceleration in our model is considerably higher than the Newtonian gravity such that the relative velocities amongst overdensities can be also much higher. We consider an alternative scenario in which a small mass protogalaxy can receive more than enough angular momentum and radial velocity towards its own mass centre from a void such that it stops growing. In this case, if an overdensity stops growing at $\delta_i$ and the mass shell at $r_i$ contracts by a factor $n$ to $r$, the rotational velocity is amplified by a factor dependent  on both $n$ and $1/\sqrt{\delta_i}$. For small $\delta_i$, this amplication can be significant. One expects that the smaller the protogalaxy mass, the smaller the angular momentum that is required to stop the overdensity growth, the smaller the $\delta_i$ (the higher the redshift $z_i$), the more severe a system's velocity dispersions will deviate from the BTFR of the massive galaxies. For Coma Ber., from a baseline overdensity of CMB, one can obtain observed rotational velocity within realistic $\delta_i$. For Segue 1 which has the smallest mass and most severe deviation from the massive galaxies' BTFR, we find that a slightly smaller baseline overdensity or a slightly excessive angular momentum transfer from a void than the just right amount is preferred. Given appropriate total energy, the orbit of a particle on a mass shell corresponding to the half light radius is bounded from above and also from below for some $r>0$. The resulting maximum uniform matter density could restrict the duration when the hydrogen density is favourable to star formation.

\section*{Data availability statement}
Data Sharing not applicable to this article as no datasets were generated or analysed during the current study.
\section*{Competing Interests}
The authors have no conflicts of interest to declare that are relevant to the content of this article.

\section*{Acknowledgements}
We thank Dr. Indranil Banik for suggesting this problem.

\end{document}